\documentclass[journal]{article}
\pdfoutput=1

\usepackage{graphicx}
\usepackage[table]{xcolor}
\usepackage{multicol}
\usepackage{url}

\author{
    C\'{e}sar Bernardini$^{1}$, Thomas Silverston$^{2}$, Athanasios Vasilakos$^{3}$\\
    $^{1}$University of Innsbruck, 6020 Innsbruck, Austria\\
    $^{2}$University of Tokyo, JFLI CNRS UMI 3527, Tokyo, Japan\\
    $^{3}$Lulea University of Technology, SE-838 73 Froson, Sweden
}

\begin{document}

\title{Caching Strategies for Information Centric Networking: Opportunities and Challenges}

\maketitle

\begin{abstract}
Internet usage has shifted from host-centric end-to-end communication to a content-centric approach mainly used for content delivery.
Information Centric Networking (ICN) was proposed as a promising novel content delivery architecture.
ICN includes in-network caching features at every node which has a major impact on content retrieval.
For instance, the ICN efficiency depends drastically on the management of the caching resources.
The management of caching resources is achieved with caching strategies.
Caching strategies decide what, where and when content is stored in the network.
In this paper, we revisit the recent technical contributions on caching strategies, we compare them and discuss opportunities and challenges for improving content delivery in the upcoming years.
\end{abstract}



\section{Introduction}

Internet usage has shifted from host-centric end-to-end communication to a content-centric approach mainly used for content delivery.
Worldwide, video consumption already represents 66\% of all Internet traffic~\cite{Bernardini:2013:MPC}.
%
Although content delivery represents such a large percentage of Internet traffic,
the paradigm of the current Internet has not been built for content delivery.
Unlike traditional broadcast which sends one title to million of people across the network at one time,
the Internet transmits same videos many times over.
In fact, 10\% of the transmitted content represents already 90\% of the Internet traffic.
If the online video consumption grows as expected,
the congestion in the Internet will get out of control and new solutions will be required to maintain an acceptable quality of service.

To address the problem, Information Centric Networks (ICN) were proposed.
ICN is a novel Internet architecture designed for content delivery.
Instead of leading the Internet protocol with an end-to-end communication protocol,
ICN switches to a content-centric approach where every content is named.
Multiple ICN architectures were proposed such as Content Centric Networking (CCN), SAIL NetInf and FP-7 PURSUIT.
Among all these new architectures, CCN has attracted most of the attraction of the community due to three reasons:
In-network caching features at every node, coupled name resolution and data forwarding and a unified naming scheme.
From these features, in-network caching impacts directly on the content delivery efficiency and it is the object of this study.

Despite the large caching literature already existing,
the premises of a CCN architecture makes its study challenging.
The in-network caching features at every node becomes CCN into a network of caches.
Internet has never handled caches at such a large scale, caches were located at fixed locations and now caches are placed everywhere.
CCN stores content at chunks of content at a fine-granularity, in contrast with traditional architecture were complete objects were stored.
CCN routers must deal with large cache sizes and a catalog ranging for all the content from the Internet.

The CCN efficiency depends drastically on the performance of its caching features.
We revisit management policies used to improve the performance of a CCN network.
These management policies are called caching strategies.
Caching strategies decide what, where and when content is stored in the network.
The contribution of this paper are twofold:

\begin{itemize}
{\item First, we present the principles for building caching strategies and revisit the solutions proposed by the literature.
}

{\item Second, we present three potential scenarios for deploying ICN where caching strategies' gains are compared.
}

{\item Then, we discuss challenges and opportunities for the development of new caching strategies.
The challenges and opportunities are classified based on a standalone and contextualized view of the caching strategies.
First, we give an internal look at the construction of caching strategies and discuss the challenges into internal structures and promising concepts currently being neglected.
Second, CCN is finding its place within the Internet:
Thanks to Software Defined Networking (SDN), CCN will be deployed into the edges of the network.
We discuss the opportunities and challenges for new caching strategies in a novel context.
}
\end{itemize}

The rest of the article is organized as follows.
The section~\ref{sec:ccn} gives an introduction to the CCN architecture and the properties of a CCN network.
In section~\ref{sec:cachingstrategies}, we present the state of the art caching strategies for CCN.
Section~\ref{sec:challenges} discuss opportunities and challenges in the development of caching strategies for the upcoming years.
Finally, we draw some conclusions in Section~\ref{sec:conclusion}.

\section{CCN Overview}
\label{sec:ccn}

\begin{figure*}[th]
    \centering
    \includegraphics[width=1\textwidth]{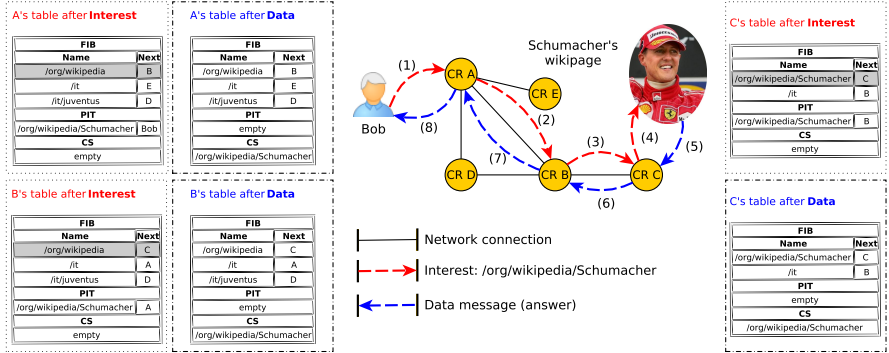}
    \vspace{-0.3cm}
    \caption{\label{fig:ccn-overview} Overview of the CCN architecture: Bob requests the Wikipedia page of Schumacher. The example depicts the process to retrieve content from a CCN network. The blocks represents the status of CRs after processing the CCN messages.}
    \vspace{-0.3cm}
\end{figure*}

Content Centric Networking (CCN) is a clean-state architecture for the future Internet.

CCN replaces the TCP/IP stack with a content centric approach.
The CCN communication paradigm is therefore led by names that represents the will of users.
The users request named content to the network and the network must provide an answer.
The CCN names are commonly known as {\it Content Names} and are structured collection of keywords.
For instance, \url{/it/Venice} and \url{/fr/Paris/EiffelTower} are two {\it Content Names} that refer to the Venice city and the Eiffel Tower.

CCN communication is based on two messages: {\it Interest} and {\it Data}.
{\it Interest} messages are generated by end-users and express the intention for a particular content.
The response arrives as a {\it Data} message.
All the messages are forwarded hop-by-hop by Content Routers (CR) in an anycast fashion.
CR maintains three data structures: {\it Forwarding Information Base} (FIB), {\it Pending Interest Table} (PIT) and {\it Content Store} (CS).
The FIB maps CCN names to output interfaces and it is used to forward {\it Interest} messages to their destination.
The PIT keeps track of the incoming interface by-which {\it Interest} messages have arrived in order to forward back the {\it Data} answer.
Finally, the CS stores information (i.e. {\it Data} messages) that has passed through the CR.
The CS serves as local cache for serving future incoming requests.

When an {\it Interest} arrives,
the CR extracts the {\it Interest} name and looks up into its CS for content stored that matches the {\it Content Name}.
If content is found,
it is sent back through the interface where it comes from.
Otherwise a longest prefix match into CR's PIT table decides whether the CR is waiting for a request of this content.
If an entry is found, the {\it Interest} is discarded.
If no entry is found, a lookup in the CR's FIB decides where the {\it Interest} is forwarded.

In the Figure~\ref{fig:ccn-overview}, we illustrate the CCN architecture.
In the example, Bob requests the wikipage of Schumacher -- an ex F1 racing driver.
Bob creates an {\it Interest} message with the {\it Content Name} \url{/org/wikipedia/Schumacher}.
The red arrows represent the path taken by the {\it Interest} message while the blue arrows the {\it Data} message's path.
The CR A looks up into their CS looking for a previously stored copy of the content.
As no match is found, the PIT table is queried.
As there is no previous {\it Interest} request for the content, there is no match in the PIT either.
As a consequence, the message is forwarded to CR B following the CR A's FIB table instructions.
The PIT tables creates a new entry for holding the path towards the requester.
The situation repeats at CR B, which forward the content through CR C.
CR C knows the exact location of Schumacher wikipage and retrieves the content.
CR C stores content into its CS and forwards back the {\it Data} message through CR B.
CR B retransmits the message to CR A which gives the Schumacher's wikipage to Bob.
Every node in the path (i.e. CR A, CR B and CR C) stores a copy of the {\it Data} message for answering future requests.

The present of caches at every CR generates interesting properties for the CCN network that are discussed in the following section.

\subsection{Properties of a CCN network}

Unlike the current Internet, CCN incorporates caching resources at every node.
Caches everywhere makes CCN a network of caches with the following properties:

\begin{itemize}
    {\item {\it Cache Transparency.}
        The cache transparency states that end-users are not aware that content is served from caches.
        End-users receive the content from a nearest-located CR while network resources are saved.
    }
    {\item {\it Cache Ubiquity.}
        Nowadays, the Internet copes with traffic demands using to Content Delivery Networks (CDN).
        CDN are proprietary proxy solutions working for particular application layer services.
        CDN disposes of caches in fixed locations such as ISP facilities and end-users are therefore served from nearest locations.
        Content placement is determined by solving an analytic model established by prior traffic demands.
        In CCN, cache are ubiquitous.
        Caches are available everywhere.
        Caching points are no longer fixed.
        The stored content is expected to adapt to the changing demands of their users.
    }

    {\item {\it Fine-Granularity of Cached Content.}
        CCN handles packets at a chunk level.
        As every single packet is named, every single chunk of content is named.
        For instance, CCN differentiates different parts of a large transmitted file while the current Internet depends on the transport and application layer.
        With CCN, the change of cache unit raises the following problem.
        In the Internet, there is a lack of unified naming scheme.
        It is difficult to collapse multiple requests for similar pieces of content and thus use the caching features.
        CDN collapses requests by looking for the URI at the application layer.
        However, if different protocols are used, the caching capabilities can not be shared.
        CCN is created with a naming schema.
        Although content is transmitted through different protocols, its name remains the same.
        Thus, cross-application cache is possible in CCN.
        This new feature permits to speed up retrieval rate and improves space utilization.
    }

\end{itemize}

The CCN properties suggest that large bulks of information will be stored in the network.
Undiscriminatory caching may have catastrophic consequences for the performance of CCN.
As such, the network must smartly select the best chunks to store, in a timely manner and foresee end-users changing traffic patterns.
A CCN network handles all these management decisions with caching strategies, a master piece in the design of CCN networks.

\section{Caching Strategies}
\label{sec:cachingstrategies}

A caching strategy is a management policy that handles temporary storage resources.
A caching strategy decides what, where and when information is stored.
In this section, we present the principles of caching strategies and then the caching strategies proposed by the literature.

\subsection{Caching Strategies Principles}

For the understanding of caching strategies, certain concepts must be first understood.

Every CCN node holds a CS table that stores content temporarily.
The CS is managed with a Replacement Policy (RP).
The RP defines the structure of the cache and the method to evict content when space is required.
The most common RP are Least Recently Used (LRU), First-In First-Out (FIFO), Last Frequently Used (LFU), Random.
RP may be grouped into the same equivalent class, i.e., different RP will exhibit the same performance in the long term.

A caching strategy targets the optimization of a metric.
The metric pursues an objective for the network.
These objectives include reduction of network traffic, tackling idleness in the network structure or decreasing end-users perceived latency.
The common-used metrics include counting the number of hits or hops and measuring the delay.
In a cache system, a hit operation occurs when a piece of content is found into a cache while a miss operation happens every time a piece of content is not found.
The most common metric is the Cache Hit, it measures the number of hit operations divided by the sum of hit and miss operations.
The number of hops is calculated averaging the number of hops required to find a content in the network.
The Frequency stands for the number of times an element is requested.
Finally, the delay metric describes the retrieval time for a piece of content within the network.

Caching strategies are also based on other concepts such as opportunistic caching and manage replica.
Opportunistic caching refers to caching the content while the {\it Data} message is routed to destination.
Managed replica consists in creating copies of the content independently of {\it Interest} and {\it Data} messages paths.

\subsection{State of the Art Caching Strategies}

We summarize the more important caching strategies developed by the CCN community:

\begin{itemize}
    {\item {\it Leave Copy Everywhere (LCE)}.
        Every time any piece of content passes by a CR, the CR stores the piece.
	    This caching strategy is the most commonly-used and the strategy by default in CCN~\cite{Zhang:2013:CachingICNSurvey}.
    }
    {\item {\it 2-Last Recently Used (2-LRU).}
        Before arriving to the CS, requests query a virtual cache that stores only the Content Name.
        In case a hit operation occurs in the virtual cache, the request is forwarded to the CS~\cite{Martina:2014:UnifiedComparison}.
    }
    {\item {\it Cache ``Less For More'' (CLFM) }~\cite{Chai:2013:CacheLess}.
        CLFM is a centrality-based caching strategy that exploits the concepts of betweenness centrality.
        With betweenness centrality, the strategy caches the content at the more central position in a path of caches.}
    {\item {\it ProbCache}~\cite{Psaras:2012:ProbCache}.
        Probabilistic In-Network CAching (ProbCache) is a probabilistic algorithm for distributed content caching along a path of caches.
        ProbCache selects the best node to cache the content in a path of caches.}
    {\item {\it Max-Gain In-Network Caching (MAGIC)}~\cite{Ren:2014:Magic}.
        This Caching strategy pretends to maximize the local gain of inserting new content in the caches,
        and to minimize the lost of replacing content.
        The maximization and minimization is calculated according to previously recorded data in the caches.}
    {\item {\it Leave Copy Down (LCD)}~\cite{Zhang:2013:CachingICNSurvey}.
        LCD replicates content only when a cache hit occurs.
        When a cache hit occurs,
        LCD copies the content into the direct neighbor towards the requester.
    }
    {\item {\it Most Popular Caching Strategy (MPC)~\cite{Bernardini:2013:MPC}.}
        MPC is a caching strategy that stores only popular content.
        MPC keeps count of the popularity of every element and replicates those pieces of content that are expected to become popular.
    }

\end{itemize}

The Table~\ref{tab:caching_strategies} summarizes the main characteristics of every caching strategy.
Most caching strategies decide whether to store information with Cache Hit metric.
Two caching strategies depend on Frequency.
However, the Frequency metric is related to the hit operations and as such can be correlated to Cache Hit.
It is also important to highlight that only two caching strategies resort to managed replication and they replicate only at one hop away.

\begin{table}[t]
    \centering
    \begin{tabular}{|l|c|c|c|}
        \hline
        Caching Strategy                & Optimized & Opp.      & Managed   \\
                                        & metric    & Caching   & Replica   \\
        \hline
        \hline
        LCE                             & Frequency & Yes       & No        \\
        2-LRU                           & Frequency & Yes       & No        \\
        Cache ``Less For More''         & Centrality& Yes       & No        \\
        MAGIC                           & Cache Hit & Yes       & No        \\

        ProbCache                       & Cache Hit & Yes       & No        \\
        LCD                             & Cache Hit & No        & Yes       \\
        MPC                             & Cache Hit & No        & Yes       \\

        \hline
    \end{tabular}
    \caption{\label{tab:caching_strategies} Summary of Caching Strategies.}
\end{table}

\section{Comparison of Caching Strategies}

Caching strategies have already been compared in a common framework in~\cite{Bernardini:2015:ComparisonCachingStrategies}.
Caching strategies must be selected according to the scenario where ICN is deployed.
However, there does not exist a list of scenarios and the best-fitted caching strategy.
In this section, we present potential scenarios for ICN networks.
For every scenario, we compare and select the best caching strategy.

We first define three scenarios where ICN networks are promising candidates:

\begin{itemize}
    {\item {\it Internet Service Provider (ISP).}
        Some content providers offer content delivery services such as video streaming or file download that are bandwidth-demanding.
        These content delivery services are sent over the ISP infrastructure and therefore its network must cope with it.
        Due to sensibility to content delivery,
        ICN appear as a promising candidate for ISP infrastructures.
    }
    {\item {\it Video On-Demand (VoD).}
        VoD services such as Youtube or Dailymotion provides already an enormous catalog of videos.
        Youtube reports that monthy 3 billions hours of video are watched and 12 million hours are uploaded.
        ICN is a good candidate for serving a network with these large numbers in content-delivery.
    }
    {\item {\it Online Social Networks(OSN).}
        The OSN usage has been empowered in the last years.
        Users share their opinion with their friends and report their experiences about services.
        As a large proportion of the exchanged content is composed of images and videos,
        the content delivery features of ICN appear make it a promising candidate.
    }
\end{itemize}

Before the analysis of caching strategies, we summarize the parameters of every scenario in the Table~\ref{tab:scenarios}.
Every scenario is composed of a popularity model, a catalog size and the average size of transmitted files.
For every parameter, we have selected the common-used values in the literature.
The popularity models follows a Heavy-tail probability function, MZipf.
The catalog sizes diverges from $10^{8}$ pieces of content for OSN and $10^{12}$ for an ISP infrastructure.
The filesizes depends on the content being transmitted.
As video files are usually larger than images or text, the VoD scenario presents an average filesize of 100MB.
Other two scenarios range from 10KB to 10MB.
The scenarios were evaluated into ISP-level topologies such as Abilene and Dtelecom.

\begin{table}[b]
    \centering
    \begin{tabular}{|l|c|c|c|}
    \hline
                            & \multicolumn{3}{c|}{Scenarios} \\
    Parameter               & ISP                   & VoD                   & OSN                  \\
    \hline
    \hline
    Chunk Size              & \multicolumn{3}{c|}{4KB}                                             \\
    Simulation Time         & \multicolumn{3}{c|}{86,400 second}                                   \\
    Topology                & \multicolumn{3}{c|}{Abilene, DTelecom}                               \\
    \hline
    \hline
    Catalog Size            & $10^{12}$             & $10^{9}$              & $10^{8}  $           \\
    Avg. Filesize           & 10KB                  & 100MB                 & 10MB                 \\
    Popularity              & MZipf($\alpha=0.65,$  & MZipf($\alpha=0.75$,   & MZipf($\alpha=1.14,$\\
    Model                   & $\beta=0.0$)          & $\beta=0.0$)          & $\beta=0.0$)         \\
    \hline
    \end{tabular}
    \caption{\label{tab:scenarios} Parameters of the Evaluation Scenarios.}
\end{table}

\begin{figure*}[th]
    \centering
    \includegraphics[width=1\textwidth]{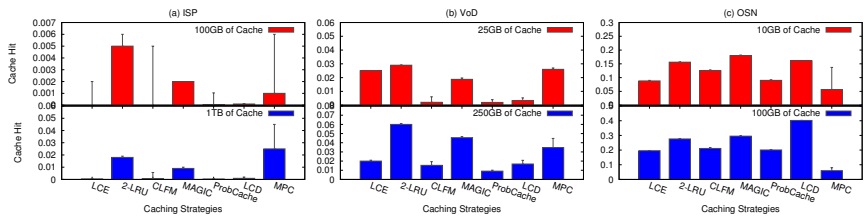}
    \vspace{-0.3cm}
    \caption{\label{fig:scenarios} Caching strategies into three ICN scenarios: (a) an ISP infrastructure, (b) a VoD network and (c) a OSN server.}
    \vspace{-0.3cm}
\end{figure*}

Figure~\ref{fig:scenarios} depicts the three scenarios.
Every scenario is shown in a separate column.
The $x-axis$ shows the caching strategies and the $y-axis$ shows the Cache Hit metric.
Every scenario is composed of two charts with different caching configuration for the ICN network.

We first evaluate the caching strategies for an ISP infrastructure in the Figure~\ref{fig:scenarios}(a).
The in-network caching configuration includes two alternatives with 100GB and 1TB of caches per node.
A glipse into the results shows that there exists only two caching strategies that appears as good candidates: 2-LRU and MPC.
In the 100GB scenario, 2-LRU reachs a 0.5\% of caching performance and it is enough to overpass all the other caching strategies.
In the 1TB scenario, MPC surpasses all the other caching strategies with 2.7\% of Cache Hit ratio.
The standard deviation must not be neglected with MPC: 
although MPC gets better results, it is unstable and may cause peaks of traffic in the rest of the network.
2-LRU is therefore the most stable and promising candidates for ISP infrastructures.

Altough Cache Hit values between 0.5\% and 2.7\% seem to be smaller,
the ISP traffic would get an important reduction.
The Telecom Italia Group transfers daily around 8.6PB~\footnote{\url{http://www.telecomitalia.com/content/dam/telecomitalia/en/archive/documents/investors/Annual_Reports/2014/Annual-report2014.pdf}}.
With an ICN network with only 1TB using 2-LRU or MPC, an ISP can save up to 232.2TB of traffic daily.
Otherwise only 100GB (i.e. a normal hard-drive) economizes up to 43TB per day.

The second scenario refers to a network of Video On-Demand services.
In the scenario, end-users request videos from a Youtube-like catalog.
The Figure~\ref{fig:scenarios}(b) represents the Cache Hit performance by deploying a ICN network with two configurations:
(1) an ICN network with 25GB of Cache and (2) another ICN with 250GB of caches.
In (1) and (2), 2-LRU overtakes the other caching strategies.
With 25GB of caches, 2-LRU reduces by 2.9\% the traffic while it reaches 6\% of reduction with 250GB of caches.
Everyday 100 million hours of video are watched in Youtube.
Youtube recommends encoding videos with H.264 and 2.5Mbps with a quality of 480p and a resolution of 854x480~\footnote{\url{https://support.google.com/youtube/answer/1722171}}.
Youtube transfers therefore 11PB per day.
As a consequence if an ICN network will help Youtube to reduce 6\% of the daily traffic, it will be translated into 676TB of traffic.

Finally, one of the most promising scenarios for ICN is analyzed: OSNs.
The Figure~\ref{fig:scenarios}(c) depicts the Cache Hit performance by deploying an ICN network in a OSN server.
Two configuration of caches are evaluated: 10GB and 100GB of caches per node.
A first look in the results shows that the Cache Hit performance is better than in the other scenarios.
This is due to the consumption pattern of OSNs:
popular publications are massively distributed while others are ignored.
The pattern produces therefore a stress into the head of the popularity model, creating a subset of super popular elements.
From the listed caching strategies, MAGIC and LCD are the most promising candidates.
In the first configuration, MAGIC achieves to save up to 18\% while LCD economizes 40\% of the traffic.
With only 100GB of caches, an ICN network may absorb 40\% of the traffic.
Clearly, showing that ICN is a promising candidate for an OSN network.

The presented scenarios give an insight into the efficiency and potential gains of ICN.
For example, the VoD network scenario has shown that a Youtube-like catalog may reduce 676TB of traffic daily by deploying an ICN network.
Furthermore, the scenarios were also useful to reduce the subset of caching strategies:
from seven caching strategies of the literature, we have reduced the subset to only four: 2-LRU, MPC, MAGIC and LCD.

\section{Opportunities and Challenges}
\label{sec:challenges}

In this section, we give an insight on opportunities and challenges for research on caching strategies.
We classify the opportunities and challenges based on an individual and global view of the caching strategies.
First, we focus on internal aspects of caching strategies that are promising for the future.
Second, as CCN networks are finding their place in the Internet, we address caching strategies in this new context.

\subsection{Internal Evolution of Caching Strategies}

Most of the caching strategies have considered the CCN nodes as a standalone CS.
However, the internal structure of a CCN node may help to improve the performance of the network.
In this section, we discuss certain premises such as nodes' internal structure or metrics that may generate opportunities in the near-future for novel caching strategies.

The efficiency of a CS depends on the ability to cope with large caches that operate at line speed.
The memory technologies that are suitable for line-speed requirements are costly and limited in space
while cheap memories provides high access latency and more storage space.
Rossini et al.~\cite{Rossini:2014:MultiTerabyteICNRouter} propose to manage the CS with both memory technologies, a small but fast cache memory and large but slow memory.
By studying the arrival probability of content, the CS can predict future requests and move content towards the fastest piece of memory.
Most of the proposed caching strategies neglect the internal structure of the CS and consider it as a plain memory.
There exists a challenge at building caching strategies that exploits the hierarchical structure of the CS.
While storing content, caching strategies must decide either to store in the fastest or the slowest memory and from which node.
Not only to migrate content locally, but also with neighbor CSs.

To route and forward content, every CCN node holds three tables: PIT, FIB and CS.
From the three tables, caching strategies have considered only one of them: the CS.
However, the other two tables, FIB and PIT, have a direct impact into caching results.
The FIB is used to forward requests and the PIT table to employ collapse forwarding.
Collapse forwarding is a cache systems' technique which aggregates similar requests from several users into single requests.
While a content is being downloaded,
a PIT entry may be responsible for delaying the request in order to collapse it with another.
Next generation caching strategies are expected to consider locally not only the CS but all the other tables.
Dehghan et al.\cite{Dehghan:2015:AnalysisCachesPIT} propose an analytical model to predict cache behavior considering the PIT table.
The model highlights the impact of the PIT table into the performance of caches and it is essential to learn from this experience.
Caching strategies must consider the effects of the PIT table in order to serve upcoming requests.

Most of the caching strategies have been assessed focusing on the Cache Hit metric.
These metrics are commonly referred as network-centric performances.
However, users are mostly interested in user-centric performances such as perceived end user delay, network congestion and link capacity constraints.
Badov et al.\cite{Badov:2014:CongestionAwareCaching} study caching strategies and forwarding mechanisms that minimize the user-perceived delay.
The congestion-aware caching strategy improves the user-perceived download time with regards to current caching strategies.
As a consequence, congestion-awareness caching strategies appear as promising approach to improve the user experience.
As the user experience motivates the change for a CCN network,
user-centric metrics seem to fit as better thermometers to measure the quality of a caching strategy.

\subsection{New Caching Strategies for a new CCN context}

CCN is finding its place within the Internet.
Technical advances suggest that CCN will be implement into the edges of the network thanks to the Software Defined Technology (SDN).
In this section, we address the challenges and opportunities for caching strategies into the new CCN context.

All the caching strategies were cross-compared to determine the most appropriated for a CCN network in~\cite{Bernardini:2015:ComparisonCachingStrategies}.
The comparison shows a broad range of scenarios for evaluating and highlight the more appropriated caching strategy for every case.
However, the most important insight is that there does not exist a killing caching strategy for every case.
Depending on the network users and request patterns, the caching strategy must be selected in a case-per-case basis.
For instance, caching strategies may be deployed for particular scenarios such as Disaster scenarios, Online Social Networks (OSN), etc.
In a disaster scenario, a CCN network must provide only a subset of sensitive information.
While in a Online Social Network, the CCN network must consider the friend-to-friend exchange patterns.

Besides the client-server retrieval scenario, one caching strategy has been deployed for OSNs.
Socially-Aware Caching Strategy (SACS) manages stored content according to the exchanged patterns of the users~\cite{Bernardini:2014:SACS}.
SACS privileges the content produced by the most influential users of the network.
This caching strategy understand how users interact and react accordingly.
As social networks alter the consumption patterns, 
more caching strategies should address interaction into OSNs.
However, it should not be limited to this particular scenario.

Software Defined Networking (SDN) is a computer networking approach that permits to manage a network through software abstractions.
SDN separates the control plane from the data plane.
The SDN controllers manages participants, organization, and configuration of the network while the data plane is in charge of the exchange of information.
Therefore, SDN technology enables the incremental deployment of ICN networks.
As networks may be deployed via software, networks can be restructured or reconfigured on-the-fly to fulfill the network needs.
Ravindran et al.\cite{Ravindran:2013:TowardsSDNICN} serves of the SDN technology to manage a CCN network as a subset of control functions.
With the SDN control plane, the CCN network may alter its forwarding tables, number of nodes and enabling real-time multimedia capabilities.
Novel caching strategies may benefit from the SDN technology:
The network may automatically replace the current caching strategy when another strategy fits better to the current network users' demand.
Metrics could be deployed to determine best-fitting strategy, avoid congestion in a network structure and models predicting gain provided by extra nodes.
%

Another interesting topic in the ICN domain is where to place caches.
CCN has been conceived as network of caching with in-network caching features everywhere.
However, it is unlikely that CCN will replace the whole Internet
and as a consequence, CCN will be implemented partially in the Internet.
Caching at the edge of the Internet seem to be the most promising solution for a CCN network~\cite{Fayazbakhsh:2013:LessPainMostGain}.
Fayazbakhsh et al.\cite{Fayazbakhsh:2013:LessPainMostGain} has shown that implementing caches in the edge of the network helps to achieve a better performance without the need of managing a enormous quantity of caches.
Imbrenda et al.\cite{Imbrenda:2014:MicroCDNApps} have proven that with only 100MB of memory allocated into the ISP facilities, say the edge of the Internet, the traffic from the ISP to the Internet can be reduced by 25\%.
The placement of caches at the edges of the network opens certain challenges for the research on caching strategies.
It is essential to determine the boundaries of the edge network in the Internet.
A CCN network may be implemented at home networks, ISP facilities or at the backbones.
It is unlikely that ISPs will replace completely their infrastructure by a CCN network.
However, there is a clear conclusion and it is that CCN will be implemented as a smaller network into the Internet.
It is clearly not the same to implement a caching strategy for less than 1000 nodes than for every Internet device into the Internet.
For every network size and location into the Internet, different caching strategies may be more fruitful.

\section{Conclusion}
\label{sec:conclusion}

CCN efficiency depends on the performance of its caching strategies.
All along this paper, we have summarized technical contributions and compare them in potential scenarios.
An ISP infrastructure, a VoD network and a OSN network have been analyzed.
From the results we have selected the four most promising caching strategies (i.e. 2-LRU, MPC, MAGIC and LCD) and we have put numbers into the potential gains of ICN.
Then, we have presented future challenges and opportunities for the development of caching strategies in CCN.
Caching strategies will evolve in two fronts in the upcoming years.

Caching strategies must revisit internal structures of a CCN node.
Caching strategies must look at other internal structures and consider more appropiate metrics.
Commonly, caching strategies regard the CCN nodes as a standalone CS without considering the PIT and FIB tables and the hierarchical composition of the CS.
Exploiting other structures may permit increase content delivery efficiency and predict future traffic demands.
With regards to metrics, caching strategies optimize network-centric metrics such as Cache Hit.
As CCN find to optimize user experience in content delivery,
user-centric metrics seem to be better alternative for building caching strategies.

Finally, caching strategies must be deployed in a novel context.
CCN networks are more likely to be deployed in the edge of the networks rather than replacing the whole Internet.
The SDN technology will permit the incremental deployment of CCN networks.
As such the CCN networks will have a short-term life with a pre-defined target such as absorbing traffic or furnishing services in a disaster scenario.

\bibliographystyle{IEEEtran}
\bibliography{references}

\end{document}